\documentclass[a4paper,aps,preprintnumbers,showpacs,twocolumn,superscriptaddress,nofootinbib,amsmath,amssymb]{revtex4}
\usepackage[dvips]{graphics}
\def\imo{i}
\def\re#1{Re(#1)}
\def\im#1{Im(#1)}

\begin{document}
\title{Quasinormal modes of massive fermions in Kerr spacetime:\\Long-lived modes and the fine structure}
\author{Roman A. Konoplya}\email{konoplya_roma@yahoo.com}
\affiliation{Theoretical Astrophysics, Eberhard-Karls University of T\"ubingen, T\"ubingen 72076, Germany}
\affiliation{Peoples Friendship University of Russia (RUDN University), 6 Miklukho-Maklaya Street, Moscow 117198, Russian Federation}
\affiliation{Institute of Physics and Research Centre of Theoretical Physics and Astrophysics, Faculty of Philosophy and Science, Silesian University in Opava, CZ-746 01 Opava, Czech Republic}
\author{Alexander Zhidenko}\email{olexandr.zhydenko@ufabc.edu.br}
\affiliation{Centro de Matem\'atica, Computa\c{c}\~ao e Cogni\c{c}\~ao, Universidade Federal do ABC (UFABC), Rua Aboli\c{c}\~ao, CEP: 09210-180, Santo Andr\'e, SP, Brazil}
\affiliation{Institute of Physics and Research Centre of Theoretical Physics and Astrophysics, Faculty of Philosophy and Science, Silesian University in Opava, CZ-746 01 Opava, Czech Republic}

\begin{abstract}
Quasinormal modes of a massive Dirac field were calculated for various static black hole backgrounds with the help of the WKB formula. These estimations, however, are rough and valid only for very small values of $\mu M$, where $M$ and $\mu$ are mass of the black hole and field respectively. Thus, no accurate calculations of massive Dirac modes are known even for the Schwarzschild black hole and this is all the more so for the  Kerr solution. Here we calculate quasinormal modes of a massive Dirac field in the Kerr background. We have shown that the infinitely long-lived quasinormal modes (quasiresonances), which exist for boson fields, appear also in the fermions' quasinormal spectrum. Two chiralities of massive fermions lead to an additional ``fine structure'' in the spectrum. We discuss the effect of this fine structure on the behavior of quasiresonances and the stability. The analysis is also extended to a charged massive field in the Kerr-Newman background.
\end{abstract}
\pacs{04.30.Nk,04.70.Bw}
\maketitle

\section{Introduction}

Recent observations of gravitational waves from, apparently, a merger of two black holes \cite{TheLIGOScientific:2016src} confirmed existence of gravitational waves, predicted a century ago by general relativity. Although the current large uncertainty in measurement of the angular momentum and mass of the resultant black hole does not allow discarding of alternative theories of gravity \cite{Konoplya:2016pmh}, the Einstein theory is fully consistent with the experimental data, and the Kerr solution \cite{Kerr:1963ud} of the Einstein equations is the most celebrated model for a rotating, axially symmetric black hole. Observationally, the most important phase of the radiation of gravitational waves is described in terms of the proper oscillation frequencies of the black hole, called \emph{quasinormal modes} \cite{reviews}.

The literature devoted to calculations of quasinormal modes of gravitational and test fields in the vicinity of a Kerr black hole is enormous by now \cite{reviews}. Nevertheless, there is an obvious gap in the study of quasinormal modes of a Dirac field in the Kerr background. Chandrasekhar  studied general properties of a massive Dirac field equation in the Kerr background \cite{Chandrasekhar:1976ap} a long time ago. However, since that time the fermionic quasinormal spectra were extensively studied either for massive fermions in the static (for example, Schwarzschild or Reissner-Nodrstr\"om) background or for massless fermions in the Kerr(-Newman) background \cite{Cho:2003qe,DiracQNMs,He:2006jv}, so that both factors (mass of the fermion field and black hole rotation) have never been considered together. Moreover, the quasinormal frequencies for static backgrounds \cite{Cho:2003qe} were found with the help of WKB formula of \cite{Iyer:1986np}, which is valid, and still approximate, only for very small values of $\mu M$. Thus, such an issue, as, for example, the existence or absence of quasiresonances\footnote{Notice that quasiresonances should not be confused with the ``quasibound states'' \cite{Rosa:2011my}, because the latter corresponds to the bound state boundary conditions (vanishing of the wave function at the event horizon and infinity), while quasiresonances are part of the spectrum of quasinormal modes.} (arbitrarily slowly damped modes) for a massive Dirac field has not been confirmed so far even for the Schwarzschild black hole. One of the reasons for avoiding such a straightforward generalization to the massive case was purely technical: angular part of the perturbation equations becomes much more complex \cite{Dolan:2009kj,angulareq,Dolan:2015eua}, leading to additional coupling terms between mass of the field $\mu$ and the rotation parameter $a$. The bound states of a massive Dirac field, which corresponds to decaying boundary conditions at both boundaries, were studied in detail in \cite{Dolan:2015eua,Huang:2017nho}.

At the same time the compact body of mass $M$ interacting with a quantum mass of a mass $\mu$ is characterized by a dimensionless parameter
\begin{equation}
\frac{\mu M}{m_{Pl}^2} = \frac{G M \mu}{\hbar c} \sim \frac{r_+}{ \lambda_{C}},
\end{equation}
where $r_+$ is the horizon radius of the compact body and $\lambda_{C}$ is the Compton wavelength of the quantum field. This way astrophysical black holes are always within the regime of  $\mu M \gg 1$, while $\mu M \lessapprox 1$ and $\mu M \ll 1$ may also include primordial and miniature black holes. From now and on we shall take $G=\hbar = c =1$.

Even though the massive fields are short ranged (so that their quasinormal spectra are unlikely to be detected by a distant observer), the study of massive fields in the vicinity of rotating black holes may be intrinsically connected to a number of other interesting problems. Thus, quasinormal modes of a massive scalar and vector fields in the Schwarzschild and Kerr backgrounds have modes with arbitrarily small damping rates, which are called \emph{quasiresonances} \cite{Ohashi:2004wr,Konoplya:2004wg,Konoplya:2005hr,quasiresonances}. It was shown in \cite{Konoplya:2004wg,Konoplya:2005hr} that for boson fields these long-lived modes do not go over into the growing modes (what would trigger an instability).

Another interesting aspect is related to the fact that an effective massive term can represent self-interaction correction to the massless field or describe the influence of a magnetic field \cite{Konoplya:2007yy}. Recently it was shown that quasinormal modes play the role of a multimode squeezer that can generate particles \cite{Su:2017fcm} and this effect is the stronger, the longer lived modes are, so that the arbitrary long-lived quasiresonances may gain now a different interesting aspect for investigation. Finally, the dominant, massive Kaluza-Klein modes appear in the scenarios with compact extra dimensions \cite{Seahra:2004fg}.

In the regime of the near extremal rotation, the quasinormal spectrum of gravitational perturbations bifurcates \cite{Yang:2012pj}, leading to the so-called \emph{Zero Damped Mode} \cite{Yang:2012pj,ZDMs} -- another way to have a long-lived mode in the spectrum. The quasinormal modes of the exactly extremal Kerr black holes have been considered in \cite{Richartz:2017qep,Burko:2017eky}. In \cite{Burko:2017eky} it was shown that the extremal state is linearly stable. However, appearing of the Zero Damped Mode in the limit of extremal rotation is, possibly, a signature of ``echoes'' from the nonlinear regime of perturbations when approaching the unstable extremal state of rotation \cite{Yang:2014tla}. It is natural to expect that a similar bifurcation may take place for fermion fields and it would be interesting to understand how both types of long-lived modes (quasiresonances and Zero Damped Mode) interfere.

Having all the above motivations in mind, here we calculate quasinormal modes a massive Dirac field in the Kerr background. The techniques used here allow one to compute also the quasinormal modes of a massive charged Dirac field for the Kerr-Newman black hole, and we derive the analytic formula for the modes in the limit of large charge of the field. In addition, by comparison of the accurate quasinormal frequencies obtained by the Frobenius method with those found by the WKB formula we will show that the WKB formula found in \cite{Iyer:1986np}, even if developed to higher orders, does not provide reliable results for massive fields, unless $\mu M$ is small enough.

The paper is organized as follows. Sec.~\ref{sec:Kerr-Newman} gives the basic formulas for the Kerr-Newman background. Sec.~\ref{sec:equations} is devoted to the general covariant equation of motion for a massive Dirac field and to the separation of angular and radial variables. Sec.~\ref{sec:Frobenius} relates the numerical method, called Frobenius expansion, allowing us to solve the angular and radial master equations. Sec.~\ref{sec:QNMs} contains description of the obtained numerical data for quasinormal modes, where Schwarzschild/Reissner-Nordstr\"om, Kerr and Kerr-Newman cases are considered separately. Finally, in Sec.~\ref{sec:conclusions} we discuss and summarize the obtained results.

\section{Kerr-Newman background}\label{sec:Kerr-Newman}

As we shall study quasinormal spectrum of the Schwarzschild, Reissner-Nordstr\"om, and Kerr black holes, it is reasonable to start the analysis of the equations of motion from the Kerr-Newman solution and then go over to the particular cases. In the Boyer-Lindquist coordinates the Kerr-Newman metric has the form
\begin{eqnarray}
ds^2 &=& \frac{\Delta(r)}{\rho^2}(dt-a\sin^2\theta d\varphi)^2-\rho^2
\left(\frac{dr^2}{\Delta(r)}+d\theta^2\right)\\\nonumber
&&-\frac{\sin^2\theta}{\rho^2}
(adt-(r^2+a^2)d\varphi)^2,
\end{eqnarray}
where
\begin{eqnarray}\nonumber
\Delta(r)&=&(r^2+a^2)-2Mr+Q^2,\\\nonumber
\rho^2&=&r^2+a^2\cos^2\theta.
\end{eqnarray}
Here $Q$ and $M$ are black hole's charge and mass respectively, $a$ is the angular momentum per unit mass. The electromagnetic background of the black hole is given by the four-vector potential
\begin{equation}
A_{\mu}dx^{\mu}
   =-\frac{Qr}{\rho^2}(dt-a\sin^2\theta d\varphi).
\end{equation}
We shall be using the following three parameters: the event horizon $r_+$, the inner horizon $r_-$, and the rotation parameter $a$,
$$0\leq a^2/r_+\leq r_-\leq r_+.$$
The black hole's mass and charge are then
$$2M=r_++r_-,\qquad Q^2=r_+r_--a^2.$$

\section{Equations of motion for a massive charged Dirac field}\label{sec:equations}

A massive charged Dirac field obeys the equations \cite{Page:1976jj}
\begin{equation}\label{Dirac}
\begin{array}{rcl}
\sqrt{2}(\nabla_{A\dot{B}}+\imo eA_{A\dot{B}})P^A+\imo\mu\overline{Q}_{\dot{B}}&=&0,\\
\sqrt{2}(\nabla_{A\dot{B}}-\imo eA_{A\dot{B}})Q^A+\imo\mu\overline{P}_{\dot{B}}&=&0.
\end{array}
\end{equation}
where $e$ and $\mu$ are the field's charge and mass, respectively, $A_{A\dot{B}}\equiv A_{\mu}\sigma^{\mu}_{A\dot{B}}$, $\nabla_{A\dot{B}}\equiv \nabla_{\mu}\sigma^{\mu}_{A\dot{B}}$ is the covariant derivative, and, conventionally, the bar over a quantity represents the complex conjugation.

Following \cite{He:2006jv}, we take
\begin{eqnarray}\nonumber
P^0&=&\frac{1}{r-\imo a\cos\theta}e^{-\imo \omega t + \imo m \phi}R_{-\frac{1}{2}}(r)S_{-\frac{1}{2}}(\theta),\\\nonumber
P^1&=&e^{-\imo \omega t + \imo m \phi}R_{+\frac{1}{2}}(r)S_{+\frac{1}{2}}(\theta),\\\nonumber
\overline{Q}^{\dot{0}}&=&-\frac{1}{r+\imo a\cos\theta}e^{-\imo \omega t + \imo m \phi}R_{-\frac{1}{2}}(r)S_{+\frac{1}{2}}(\theta),\\\nonumber
\overline{Q}^{\dot{1}}&=&e^{-\imo \omega t + \imo m \phi}R_{+\frac{1}{2}}(r)S_{-\frac{1}{2}}(\theta),
\end{eqnarray}
and obtain the following equations for the angular part:
\begin{eqnarray}\nonumber
&&\left(\frac{\partial^2}{\partial \theta^2} + (\cot \theta + M(\theta))\frac{\partial}{\partial\theta} -2s(M(\theta)H(\theta)+H'(\theta))\right.
\\\label{angularpart}
&&-\frac{1}{2\sin^2\theta}+\frac{\cot\theta M(\theta)}{2}-H^2(\theta)+\frac{1}{4}\cot^2\theta
\\\nonumber&&\left.+\lambda^2 - \mu^2 a^2 \cos^2 \theta \right) S_s(\theta) = 0,
\end{eqnarray}
where $\lambda$ is the separation constant, $s=\pm1/2$,
$$H(\theta)=a\omega\sin\theta-\frac{m}{\sin\theta},\quad M(\theta)=\frac{a\mu\sin\theta}{a\mu\cos\theta-2s\lambda}.$$

For the radial part one can find \cite{He:2006jv}
\begin{eqnarray}\nonumber
&&\left(\Delta^{-s}(r)\frac{d}{dr}\Delta^{s+1}(r)\frac{d}{dr}+\frac{2\imo s\mu\Delta}{\lambda-2\imo s\mu r}\frac{d}{dr}+2\imo sK'(r)\right.
\\\label{radialpart}
&&+\frac{K^2(r)-\imo sK(r)\Delta'(r)}{\Delta(r)}-\frac{\mu K(r)}{\lambda-2\imo s\mu r}-\mu^2r^2-\lambda^2
\\\nonumber&&\left.+\left(s+\frac{1}{2}\right)\left(1+\frac{\imo\mu}{\lambda-2\imo s\mu r}\frac{\Delta'(r)}{2}\right)\right)R_s(r)=0,
\end{eqnarray}
where the eigenvalue $\lambda$ depends on $\omega$ and
$$K(r)=(r^2 + a^2)\omega-am - eQr.$$

Further we can use numerical methods for finding eigenvalues $\lambda$ and $\omega$.

\section{Frobenius method}\label{sec:Frobenius}

\subsection{Angular part}
We solve the angular equation (\ref{angularpart}) numerically, using the Frobenius method. Following \cite{angulareq}, we introduce a new variable
$x=\dfrac{1+\epsilon\cos\theta}{2}$,
where $\epsilon=\dfrac{m}{|m|}$.
Then the angular parts $S_{\pm1/2}$ can be expanded as
\begin{eqnarray}\nonumber
S_{1/2}(x)&=&x^{p/2}(1-x)^{(p+1)/2}\sum_{n=0}^{\infty}C_n Q^+_n(x)\\\label{angularexpansion}
S_{-1/2}(x)&=&\pm(1-x)^{p/2}x^{(p+1)/2}\times\\\nonumber&&\times\sum_{n=0}^{\infty}(-1)^n\frac{n+p+1}{p+1}C_n Q^-_n(x),
\end{eqnarray}
where $p\equiv|m|-1/2$ and
$$Q^{\pm}_n(x)\equiv{}_2F_1\left(-n,n+2p+2,p+\frac{3\mp1}{2},x\right)$$
are the Gaussian hypergeometric functions.

\begin{figure*}
\resizebox{\linewidth}{!}{\includegraphics*{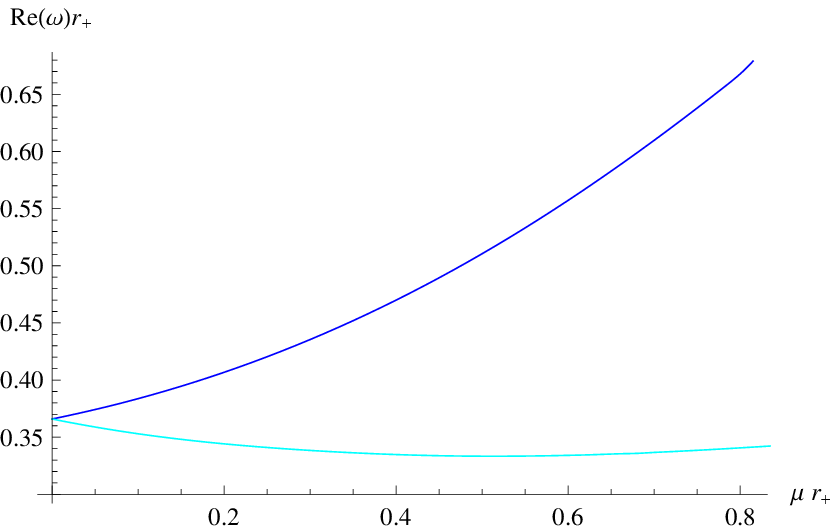}\includegraphics*{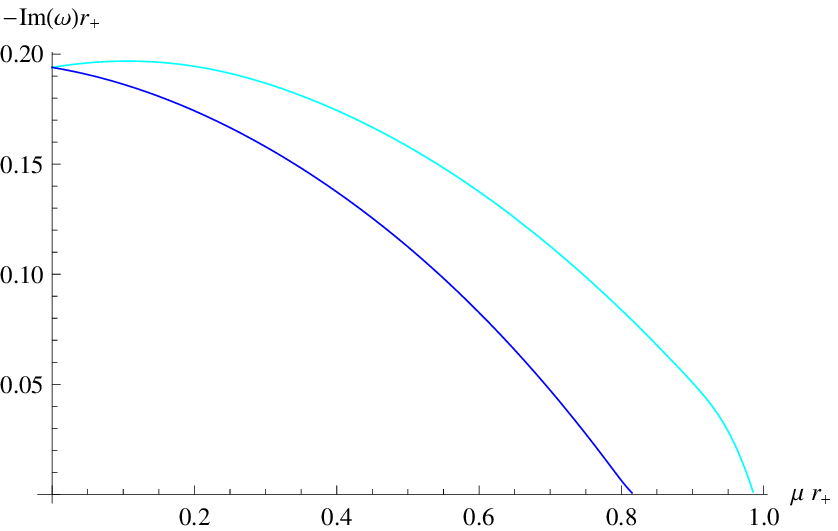}}
\caption{$\re{\omega}$ (left) and $\im{\omega}$ (right) for the fundamental ($n=0$, $\ell =1/2$) quasinormal mode of the neutral, massive Dirac field in the Schwarzschild background: $\sigma=1/2$ (blue) and $\sigma=-1/2$ (cyan). Positive value of the chirality corresponds to higher oscillation frequency and slower decay.}\label{S}
\end{figure*}

The two possible signs in (\ref{angularexpansion}) correspond to the different chiralities of the massive Dirac field, which we enumerate by a half-integer number\footnote{In the notations of \cite{angulareq}, $\sigma=\frac{1}{2}\epsilon_\lambda$.} $\sigma=\pm\frac{1}{2}$. This leads to a fine structure in the quasinormal spectrum of a massive Dirac field (see Fig.~\ref{S}).

The coefficients $C_0,C_1,C_2\ldots$ satisfy the three-term recurrence relation \cite{angulareq}
\begin{equation}
\alpha_n C_{n+1}+\beta_n C_n+\gamma_n C_{n-1}=0,\quad C_{-1}=0,
\end{equation}
with the coefficients
\begin{eqnarray}
\alpha_n&=&\epsilon\frac{n+1}{2n+2p+3}\left(\nu a\mu(-1)^n+a\omega\right),\\
\beta_n&=&(n+p+1)-\nu\lambda(-1)^n\\\nonumber&&+\epsilon\frac{(2p+1)(\nu a\mu(-1)^n-2a\omega(n+p+1))}{(2n+2p+1)(2n+2p+3)},\\
\gamma_n&=&\epsilon\frac{n+2p+1}{2n+2p+1}\left(\nu a\mu(-1)^n-a\omega\right),
\end{eqnarray}
where $\nu=\pm1$, corresponding to the two possible signs of $\sigma$.

Next, for any given $\omega$, we can calculate $\lambda$ as the most stable root of the equation with an infinite continued fraction \cite{angulareq}
\begin{eqnarray}\label{continued_fraction}
\beta_N-\dfrac{\alpha_{N-1}\gamma_{N}}{\beta_{N-1}
-\dfrac{\alpha_{N-2}\gamma_{N-1}}{\beta_{N-2}-\alpha_{N-3}\gamma_{N-2}/\ldots}}=\\\nonumber
\dfrac{\alpha_N\gamma_{N+1}}{\beta_{N+1}-\dfrac{\alpha_{N+1}\gamma_{N+2}}{\beta_{N+2}-\alpha_{N+2}\gamma_{N+3}/\ldots}}.
\end{eqnarray}
The infinite number of roots are enumerated by a nonnegative integer $N=0,1,2,3\ldots$.

For the nonrotating case $\alpha_n=\gamma_n=0$ and the equation (\ref{continued_fraction}) is reduced to \cite{angulareq}
$$\beta_N=(N+p+1)-\nu\lambda(-1)^N=0,$$
implying that $\lambda$ is a positive or negative integer number, such that $|\lambda|>|m|$,
\begin{equation}\label{nonrot}
\lambda=(-1)^N(N+p+1)/\nu=(2\ell+1)\sigma,
\end{equation}
where we have introduced a half-integer positive multipole number $\ell=|m|+N$.

Finally, recalling that $p=|m|-1/2$, we can find the value of $\nu$ for given $\ell$, $m$, and $\sigma$ as follows
$$\nu=\frac{(-1)^{\ell-|m|}}{2\sigma}.$$

For the nonzero rotation ($a>0$), in order to find $\lambda$, which corresponds to the given values of $\ell$, $m$, and $\sigma$, we slowly increase $a$, finding for each step the closest solution for the eigenvalue $\lambda$, starting from the exact value for the nonrotating case (\ref{nonrot}).

\subsection{Radial part}
We solve the master equation for the radial part (\ref{radialpart}) at the quasinormal boundary conditions. For  asymptotically flat black holes, which we consider here, these boundary conditions correspond to the purely ingoing wave at the event horizon and the purely outgoing wave at infinity (see \cite{reviews} for reviews).
Since $R_{\pm\frac{1}{2}}$ can be expressed through $R_{\mp\frac{1}{2}}$ and its first derivative (see Eqs. (2.9) and (2.10) of \cite{He:2006jv}), the equations for $s=\pm1/2$ are isospectral. Although for the simplicity we use $s=-1/2$, we have checked that the calculations for $s=1/2$ lead to the same quasinormal modes within the chosen numerical precision.

We solve Eq.~(\ref{radialpart}) using the Frobenius method (see \cite{Zhidenko:2009zx} for details). The solution can be written in the following form
\begin{eqnarray}\label{Frobenius}
R_s(r)&=&e^{\imo\Omega r}(r-r_-)^{\alpha}\left(\frac{r-r_+}{r-r_-}\right)^{-s-\imo K(r_+)/\Delta'(r_+)}\\\nonumber&&
\times\sum_{k=0}^{\infty}a_k\left(\frac{r-r_+}{r-r_-}\right)^k, \qquad \Omega=\sqrt{\omega^2-\mu^2},
\end{eqnarray}
where $\re{\Omega}$ and $\re{\omega}$ have the same sign,
$$\alpha=\frac{\imo eQ\omega -(r_++r_-)(\omega^2-\mu^2/2)}{\Omega}-\left(s+\frac{1}{2}\right)$$
is fixed in such a way that the series is convergent if and only if $\omega$ is a quasinormal mode.

Substituting (\ref{Frobenius}) into (\ref{radialpart}), we find the four-term recurrence relation for the coefficients $a_k$,
\begin{equation}\label{fourterm}
c_{0,n}a_n+c_{1,n}a_{n-1}+c_{2,n}a_{n-2}+c_{3,n}a_{n-3}=0.
\end{equation}
By performing the Gaussian elimination, for each $n$ from (\ref{fourterm}) we obtain the three-term relation,
which allows us to reduce the problem of calculation of $\omega$ to solving a nonalgebraic equation with infinite continued fraction \cite{Leaver:1985ax}.

\begin{figure*}
\resizebox{\linewidth}{!}{\includegraphics*{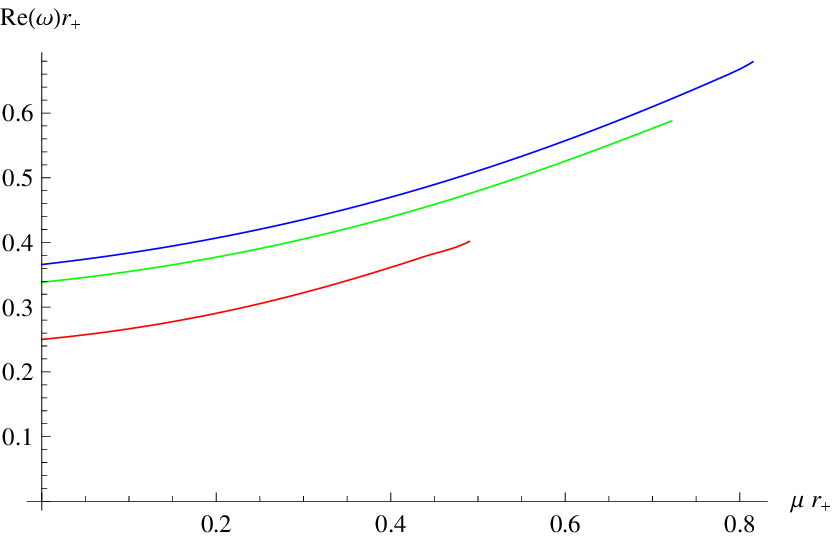}\includegraphics*{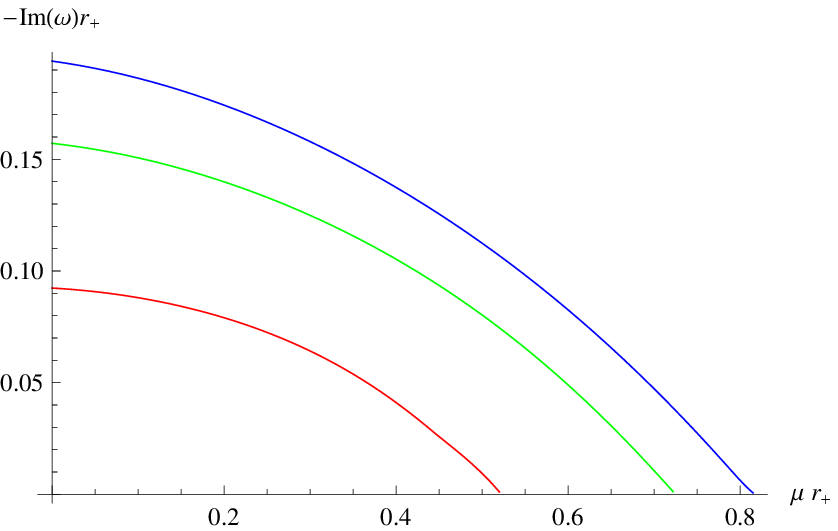}}
\caption{$\re{\omega}$ (left) and $\im{\omega}$ (right) for the fundamental quasinormal mode ($\ell =1/2$, $n=0$) of the neutral Dirac field ($\sigma=1/2$) as a function of the field mass ($\mu r_+$) in the Reissner-Nordstr\"om black hole background: $Q =0$ (top, blue), $Q=0.5r_+$ (green), $Q=0.95r_+$ (bottom, red).}\label{RN1}
\end{figure*}

\section{Quasinormal modes}\label{sec:QNMs}
\subsection{Symmetries}
Solutions to the master wave equation possess a number of symmetries which must be mentioned here in order to avoid repeated calculations of the same modes.
For given parameters of the black hole ($M$, $a$, $Q$) and the Dirac field ($\mu$, $e$) the quasinormal frequencies are enumerated by the half-integer positive multipole number,
$$\ell=\frac{1}{2}, \frac{3}{2}, \frac{5}{2}, \ldots,$$
azimuthal number,
$$m=-\ell,-\ell+1,\ldots,\ell-1,\ell,$$
chirality,
$$\sigma=\pm \frac{1}{2},$$
and the overtone number $$n=0,1,2,3\ldots. $$

From the radial equation (\ref{radialpart}) it is easy to see that the simultaneous changing of signs of $\mu$ and $\sigma$ (the latter changes sign of $\lambda$)
\begin{equation}\label{symmetry1}
\mu \rightarrow -\mu, \quad \sigma \rightarrow -\sigma,
\end{equation}
leads to the same wave equation (\ref{radialpart}) for the nonrotating case. The symmetry (\ref{symmetry1}) takes place also for the rotating case, though it is less straightforward to show.

In addition, as in the massless case, for a massive field there is the following symmetry
\begin{equation}
m\to-m, \quad e\to-e, \quad \re{\omega}\to-\re{\omega}.
\end{equation}

That is why here we will present the data only for $\mu\geq0$ and $\re{\omega}\geq0$.

\subsection{Massive neutral Dirac field in the Schwarzschild and Reissner-Nordstr\"om backgrounds}

While quasinormal modes of a massless Dirac field were accurately calculated with different converging procedures, the massive case was considered only with the WKB method \cite{Cho:2003qe} developed in \cite{Iyer:1986np}. Strictly speaking, the WKB method \cite{Iyer:1986np} cannot be applied to the massive case, as the effective potential allows for another local minimum far from black hole, so that the problem has now three turning points. Moreover, if for tiny values of the mass $\mu$ the subscattering around the second minimum could be neglected, at sufficiently large $\mu$ even the main peak of the effective potential is absent. This made some authors claim that the quasinormal modes are absent for large values of $\mu$ \cite{Cho:2003qe}.

Such a WKB approach, evidently cannot detect quasiresonances accurately, so that the question whether quasiresonances exist also for the Dirac field remained open. Here, using the Frobenius method, we present accurate values for the quasinormal modes of a massive Dirac field in the Schwarzschild and Reissner-Nordstr\"om backgrounds and show that quasiresonances do exist for the spinor field as well. Indeed, from Figs.~\ref{S}~and~\ref{RN1} one can see that when $\mu$ grows, the damping rate decreases, and dependence on the field mass is roughly linear in the regime of small $\im{\omega}$. From Fig.~\ref{S} one can see that the spectrum of the Schwarzschild black hole gets an additional splitting according to the couple of chiralities $\sigma=\pm 1/2$, so that the positive chirality is characterized by a higher real oscillation frequency and longer lifetime.

Comparison with the WKB data, for example, $\omega r_+ = 0.328 - 0.208\imo$ (from Eq.~(57) of \cite{Cho:2003qe}) for $\mu r_+ =0.2$ with our value\footnote{Note that $\kappa>0$ in \cite{Cho:2003qe} corresponds to $\sigma=-1/2$ in our notations.} $\omega r_+=0.344169-0.194476\imo$ (Table~\ref{QNMs}) shows that the WKB formula is, as expected, works badly even for relatively small masses of the field, giving a relative error more than $5\%$. Thus, the WKB results for a massive Dirac field obtained in \cite{Cho:2003qe} have a significant error, which is usually of the same order as the effect and, for sufficiently large mass $\mu$, can produce even senseless results. The question whether the WKB formula developed in \cite{Iyer:1986np} can be applied effectively to the case of a massive field is discussed in details in the appendix to this paper and our conclusion is that it gives reasonable accuracy only for sufficiently large multipole numbers, small overtones and \emph{far from the quasiresonant state}.

Extrapolating the modes with the vanishing damping rates, we find the asymptotic value of mass of the field (for $\ell=1/2$, $\sigma=1/2$) at which the nondecaying state is (asymptotically) achieved:
\begin{eqnarray}\nonumber
Q =0: &\quad& \mu r_+= 0.816,\\
Q =0.5r_+: &\quad&  \mu r_+= 0.722,\\
\nonumber
Q =0.95r_+: &\quad& \mu r_+= 0.521.
\end{eqnarray}
Thus, the near extremal black hole is characterized by relatively smaller masses (in units of the horizon radius) at which quasiresonance regime is achieved. However, the corresponding asymptotic value of $\mu M=0.408$ is the smallest for the uncharged (Schwarzschild) black hole (for $\ell=1/2$, $\sigma=1/2$).

\subsection{Massive Dirac field in the Kerr background}

\begin{table*}
\begin{tabular}{|c|c|c|c|c|c|c|}
 \hline
 $n=0$ & $\ell=1/2$ & $\mu=0$ & \multicolumn{2}{c|}{$\mu r_+=0.1$}& \multicolumn{2}{c|}{$\mu r_+=0.2$}\\
 \hline
 $a/r_+$ & $m$ & $\sigma=\pm1/2$ & $\sigma=+1/2$ & $\sigma=-1/2$ & $\sigma=+1/2$ & $\sigma=-1/2$\\
 \hline
 $0$ & $\pm1/2$ & $0.365926-0.193965\imo$ & $0.383713- 0.186324\imo$ & $0.352978-0.196821\imo$ & $0.406774-0.174251\imo$ & $0.344169-0.194476\imo$ \\
 \hline
 $0.1$ & $+1/2$ & $0.378027-0.190214\imo$ & $0.395756-0.183230\imo$ & $0.364925-0.192841\imo$ & $0.416509-0.173010\imo$ & $0.355850-0.190728\imo$ \\
 $0.1$ & $-1/2$ & $0.349639-0.192560\imo$ & $0.367417-0.184361\imo$ & $0.336973-0.195523\imo$ & $0.390834-0.171328\imo$ & $0.328596-0.192803\imo$ \\
 \hline
 $0.2$ & $+1/2$ & $0.385336-0.181297\imo$ & $0.402891-0.175066\imo$ & $0.372245-0.183586\imo$ & $0.425214-0.165129\imo$ & $0.363096-0.181591\imo$ \\
 $0.2$ & $-1/2$ & $0.329989-0.186557\imo$ & $0.347744-0.177897\imo$ & $0.317696-0.189496\imo$ & $0.371595-0.163965\imo$ & $0.309883-0.186242\imo$ \\
 \hline
 $0.3$ & $+1/2$ & $0.387525-0.167723\imo$ & $0.404753-0.162332\imo$ & $0.374639-0.169586\imo$ & $0.426588-0.153571\imo$ & $0.365630-0.167625\imo$ \\
 $0.3$ & $-1/2$ & $0.307930-0.176956\imo$ & $0.325700-0.167918\imo$ & $0.296060-0.179743\imo$ & $0.350132-0.153127\imo$ & $0.288913-0.175780\imo$ \\
 \hline
 $0.4$ & $+1/2$ & $0.384559-0.150381\imo$ & $0.401273-0.145911\imo$ & $0.372094-0.151745\imo$ & $0.422491-0.138405\imo$ & $0.363449-0.149771\imo$ \\
 $0.4$ & $-1/2$ & $0.284468-0.164994\imo$ & $0.302335-0.155634\imo$ & $0.273038-0.167513\imo$ & $0.327558-0.139984\imo$ & $0.266627-0.162667\imo$ \\
 \hline
 $0.5$ & $+1/2$ & $0.376633-0.130338\imo$ & $0.392624-0.126878\imo$ & $0.364824-0.131146\imo$ & $0.413075-0.120727\imo$ & $0.356775-0.129140\imo$ \\
 $0.5$ & $-1/2$ & $0.260595-0.151903\imo$ & $0.278673-0.142237\imo$ & $0.249602-0.154059\imo$ & $0.304935-0.125676\imo$ & $0.243962-0.148158\imo$ \\
 \hline
 $0.6$ & $+1/2$ & $0.364057-0.108653\imo$ & $0.379091-0.106321\imo$ & $0.353160-0.108844\imo$ & $0.398617-0.101681\imo$ & $0.345943-0.106812\imo$ \\
 $0.6$ & $-1/2$ & $0.237222-0.138683\imo$ & $0.255628-0.128683\imo$ & $0.226654-0.140405\imo$ & $0.283189-0.111108\imo$ & $0.221793-0.133295\imo$ \\
 \hline
 $0.7$ & $+1/2$ & $0.347055-0.086247\imo$ & $0.360861-0.085247\imo$ & $0.337359-0.085731\imo$ & $0.379319-0.082414\imo$ & $0.331224-0.083689\imo$ \\
 $0.7$ & $-1/2$ & $0.215070-0.125976\imo$ & $0.233903-0.115589\imo$ & $0.204922-0.127210\imo$ & $0.263007-0.096867\imo$ & $0.200823-0.118765\imo$ \\
 \hline
 $0.8$ & $+1/2$ & $0.325366-0.063890\imo$ & $0.337611-0.064689\imo$ & $0.317254-0.062457\imo$ & $0.354989-0.064358\imo$ & $0.312501-0.060393\imo$ \\
 $0.8$ & $-1/2$ & $0.194573-0.114108\imo$ & $0.213913-0.103279\imo$ & $0.184847-0.114806\imo$ & $0.244796-0.083271\imo$ & $0.181489-0.104933\imo$ \\
 \hline
 $0.9$ & $+1/2$ & $0.296610-0.042676\imo$ & $0.307020-0.047290\imo$ & $0.290891-0.039461\imo$ & $0.325234-0.051511\imo$ & $0.288091-0.037193\imo$ \\
 $0.9$ & $-1/2$ & $0.175908-0.103227\imo$ & $0.195830-0.091903\imo$ & $0.166612-0.103339\imo$ & $0.228731-0.070476\imo$ & $0.163979-0.091973\imo$ \\
 \hline
\end{tabular}
\caption{Dominant quasinormal modes ($\omega r_+$) of a massive Dirac field in the Kerr background.}\label{QNMs}
\end{table*}

\begin{figure*}
\resizebox{\linewidth}{!}{\includegraphics*{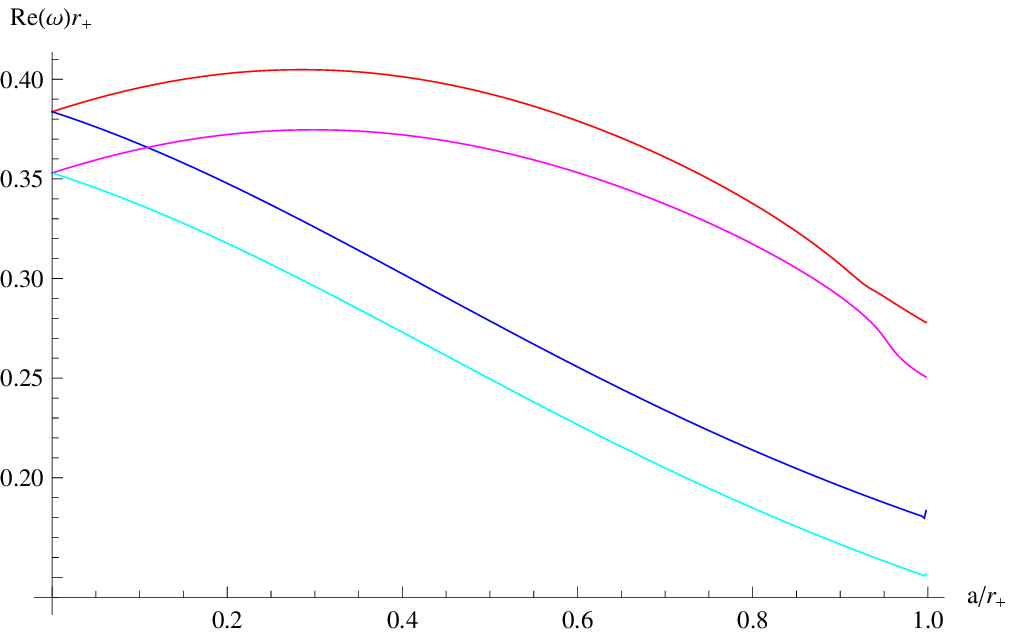}\includegraphics*{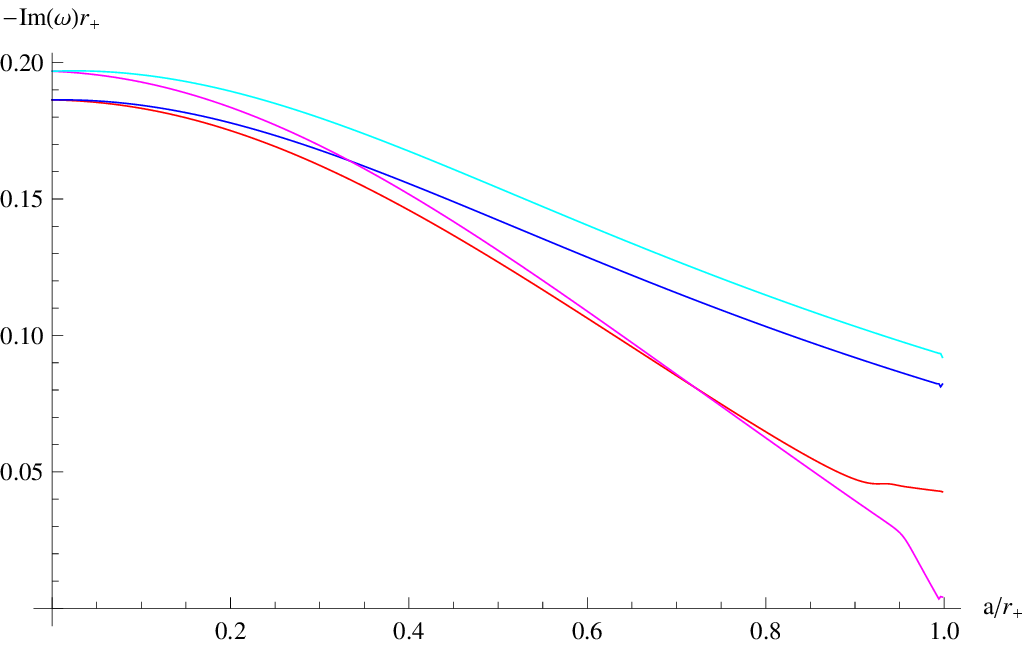}}
\caption{$\re{\omega}$ (left) and $\im{\omega}$ (right) for the fundamental quasinormal mode ($\ell=1/2$, $n=0$) of the neutral ($e=0$), massive ($\mu r_+=0.1$) Dirac field in the Kerr ($Q=0$) background: $\sigma=1/2$, $m=1/2$ (red); $\sigma=1/2$, $m=-1/2$ (blue); $\sigma=-1/2$, $m=1/2$ (magenta); $\sigma=-1/2$, $m=-1/2$ (cyan). The positive value of the chirality corresponds to higher oscillation frequency and slower decay in the nonrotating limit. The positive azimuthal value leads to higher oscillation frequency and slower decay for the rotating black hole. Yet, for the near-extremal rotation $\sigma=-1/2$, $m=1/2$ has the smallest, almost vanishing, decay rate.}\label{Kerr}
\end{figure*}

For the rotating case, not only states with opposite chirality but also states with different azimuthal numbers $m$ are nondegenerate. In table~\ref{QNMs} and Fig.~\ref{Kerr} one can see that the positive $m$ corresponds to larger values of $\re{\omega}$ for both chiralities. The positive chirality has real oscillation frequency, which is larger than that for a negative chirality and the same azimuthal number $m$. The modes with positive $m$ are longer lived than those with a negative one and the same chirality. The positive chirality modes with the same azimuthal number $m$ are longer lived than the negative chirality modes only for not very high rotation. At some $a$ which depends on the values of $\ell$ and $m$, the curves of $\sigma=1/2$ and $\sigma=-1/2$ modes intersect and the mode for negative chirality becomes longer lived. This phenomenon reflects the interplay between approaching the quasiresonance and decreasing of the decay rate for the positive values of $m$ due to extraction of rotational energy from a black hole. Rather irregular dependence of the quasinormal frequencies on $a$ in the regime of the near extremal rotation (see Fig. \ref{Kerr}) takes place also for the massless Dirac quasinormal modes \cite{Konoplya:2007zx} and for a massive charged scalar field in the Kerr-Newman background \cite{Konoplya:2013rxa}. We have checked all the numerical data against the convergence of the continued fraction.

\subsection{Massive charged Dirac field}

\begin{figure*}
\resizebox{\linewidth}{!}{\includegraphics*{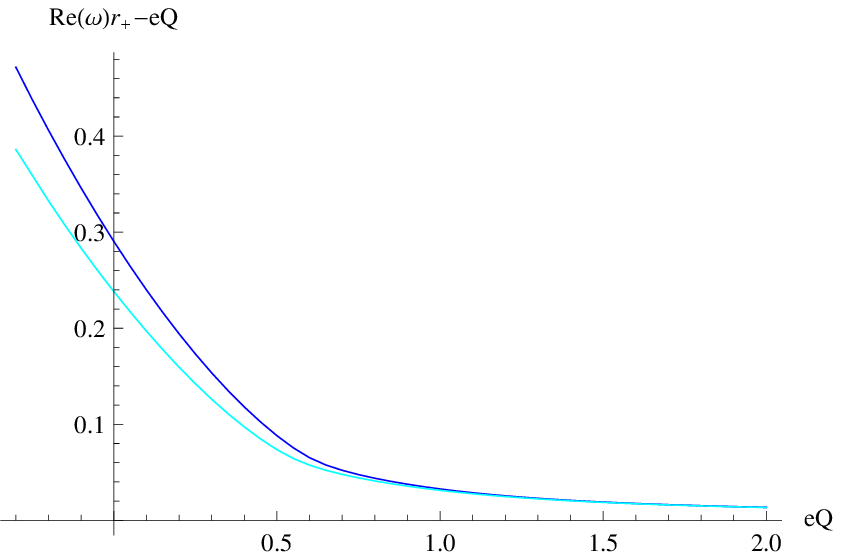}\includegraphics*{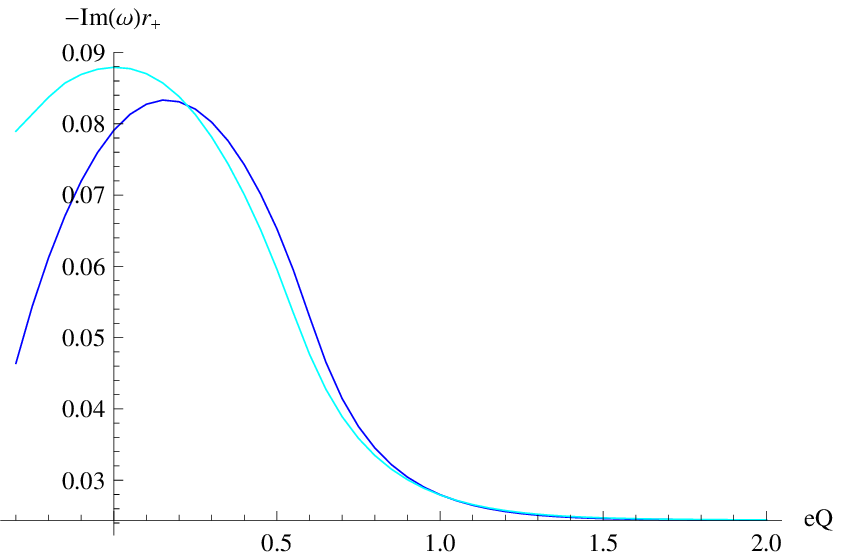}}
\caption{$\re{\omega}$ (left) and $\im{\omega}$ (right) for the fundamental quasinormal mode ($\ell=1/2$, $n=0$) of the massive ($\mu r_+=0.2$) Dirac field as a function of the field charge ($eQ$) for the Reissner-Nordstr\"om black hole ($Q=0.95r_+$): $\sigma=1/2$ (blue) and $\sigma=-1/2$ (cyan). The horizontal axis corresponds to the asymptotic value ($\omega r_+=eQ-0.024375\imo$) given by the formula (\ref{asymptotical_frequency}).}\label{eQ}
\end{figure*}

The charged Dirac field is characterized by larger (smaller) oscillation frequency for positive (negative) $eQ$ respectively (see Fig.~\ref{eQ}). As the damping rate approaches zero, it is highly sensitive to even small changes of any of the parameters  of the system in the regime of quasiresonances.
This behavior can be understood by considering dependence of the quasinormal mode on the field mass when all other parameters are fixed. In particular, for each value of $eQ$ there is some value of mass for which the quasiresonance is reached. For larger mass the corresponding mode disappears from the spectrum, and the next overtone becomes dominant. We observe that for smaller $eQ$ this threshold value of mass decreases, so that if we vary $eQ$ for a fixed mass we can approach the quasiresonance for a sufficiently small (negative) value of $eQ$.
Thus, for example, the fundamental mode for $\mu r_+= 0.49$, $Q=0.95r_+$ is $\omega r_+ = 0.4014 - 0.0108 \imo$ for the neutral field and $\omega r_+= 0.4210 - 0.0233 \imo$ for the charged field with $eQ = 1/20$, that is, even for tiny $eQ$ the damping rate is increased by a factor 2. The smaller damping rate is, the slower is the convergence of the continued fraction, making it time consuming to find frequencies which are close to the real axis.

As in the case of the scalar field \cite{Konoplya:2013rxa}, we observe that for $eQ\gg1$, $\omega={\cal O}(eQ)\gg\mu$. Then we find that $c_{1,n}={\cal O}(eQ)^2$, $c_{2,n}={\cal O}(eQ)^2$, $c_{3,n}={\cal O}(eQ)^2$, while $c_{0,n}={\cal O}(eQ)$. Therefore, the equation with the continued fraction in this regime is reduced to
\begin{equation}\label{continued_fraction_eQ}
\frac{c_{1,n}}{(eQ)^2} + {\cal O}\left(\frac{1}{eQ}\right)=0.
\end{equation}
Following \cite{Konoplya:2013rxa}, we write down for $\lambda$ when $\omega\gg\mu$,
$$\lambda^2=\lambda_0 a\omega+{\cal O}(1),$$
where $\lambda_0$ is a constant. Finally, from (\ref{continued_fraction_eQ}) after some algebra we obtain the asymptotic formula for the quasinormal modes of a charged massless Dirac field
\begin{eqnarray}\label{asymptotical_frequency}
\omega(a^2+r_+^2)=eQr_++a\left(m+\frac{r_+(r_+-r_-)}{r_+^2-a^2}\frac{\lambda_0}{4}\right)\\\nonumber-\imo(r_+-r_-)\frac{2n+1}{4}+{\cal O}\left(\frac{1}{eQ}\right)\,,
\end{eqnarray}
which has the same form as the one obtained in \cite{Konoplya:2013rxa} for the scalar field.

The formula (\ref{asymptotical_frequency}) is obtained in the regime when charges of the Dirac field and Kerr-Newman black hole are large and $\omega={\cal O}(eQ)\gg\mu$. In this regime we see that the frequencies depend on $\ell$, $\sigma$, and $\mu$ through the constant $\lambda_0$ only. In particular, for the nonrotating (Reissner-Nordstr\"om) black hole we observe degeneracy of the spectrum in this limit (see Fig.~\ref{eQ}). However, even for large $eQ$ one can reach quasiresonances by increasing the field mass. The corresponding mass has to be large enough in order to satisfy the necessary condition $\mu>\re{\omega}$ \cite{Konoplya:2004wg}.

\section{Final remarks}\label{sec:conclusions}

Although  the quasinormal modes of a massive Dirac field in the Schwarzschild background were considered in a few papers for static black holes, no such analysis for the Kerr metric was suggested. Moreover, previous results for the Schwarzschild spacetime used the WKB approach and did not allow one to analyze the regime of long-lived modes even for the Schwarzschild metric. Here with the help of the convergent Frobenius method we computed quasinormal modes of a massive charged Dirac field in the background of the Schwarzschild, Reissner-Nordstr\"om, and Kerr(-Newman) black holes. For the first time we have shown that the infinitely long-lived modes, called quasiresonances, exist not only for boson, but also for fermion fields. The massive term leads to an additional fine structure, related to the coupling between the chirality and mass of the field.

We observed that the quasiresonances do not go over into any kind of growing modes, what should be interpreted on behalf of  stability of a massive Dirac field in the Kerr background under quasinormal mode boundary conditions. However, the strict proof of stability must be done via analysis of all the possible modes of the spectrum either analytically or numerically. The latter could be fulfilled, for example, by the time-domain integration of the wave equation, which would take into consideration contributions of all overtones.

While our paper was in preparation, there appeared the work \cite{Blazquez-Salcedo:2017bld} devoted to quasinormal modes of a massive neutral Dirac field in the $D$-dimensional Schwarzschild background. Their results coincide in the four-dimensional case with ours, when the mass of the field is relatively small. Note, that $sign(\kappa)=\pm 1$  in \cite{Blazquez-Salcedo:2017bld} corresponds to $\sigma=\mp 1/2$ in our notations.

\acknowledgments{
The publication has been prepared with the support of the ``RUDN University Program 5-100''. R.~K. and A.~Z. acknowledge support of the ``Project for fostering collaboration in science, research and education'' funded by the Moravian-Silesian Region, Czech Republic and of the Research Centre for Theoretical Physics and Astrophysics, Faculty of Philosophy and Science of Sileasian University at Opava. A.~Z. was partially supported by Conselho Nacional de Desenvolvimento Cient\'ifico e Tecnol\'ogico (CNPq), Brazil. We would like to thank Sam Dolan for a most useful discussion and sharing his code for calculation of the eigenvalues for  the angular part developed in \cite{Dolan:2009kj}.}\\

\appendix

\section*{Appendix: Can Will-Schutz-Iyer formula be applied to massive fields?}

\begin{table*}
\begin{tabular}{|l|c|c|c|}
  \hline
 $\mu r_+$ & Frobenius & 3rd order WKB & 6th order WKB \\
  \hline
$ 0.1 $&$ 0.968866-0.192976 \imo $&$ 0.968000-0.193068 \imo $&$ 0.968862-0.192991 \imo $\\
   \hline
$ 0.2 $&$ 0.973608-0.191349 \imo $&$ 0.972740-0.191438 \imo $&$ 0.973603-0.191364 \imo $\\
  \hline
$ 0.3 $&$ 0.981527-0.188624 \imo $&$ 0.980658-0.188708 \imo $&$ 0.981523-0.188638 \imo $\\
   \hline
$ 0.4 $&$ 0.992653-0.184778 \imo $&$ 0.991781-0.184857 i $&$ 0.992649-0.184793 \imo $\\
   \hline
$ 0.5 $&$ 1.007024-0.179782 \imo $&$ 1.006149-0.179853 \imo $&$ 1.007020-0.179796 \imo $\\
   \hline
$ 0.6 $&$ 1.024693-0.173590 \imo $&$ 1.023813-0.173652 \imo $&$ 1.024689-0.173604 \imo $\\
   \hline
$ 0.7 $&$ 1.045724-0.166141 \imo $&$ 1.044841-0.166192 \imo $&$ 1.045720-0.166154 \imo $\\
   \hline
$ 0.8 $&$ 1.070200-0.157351 \imo $&$ 1.069311-0.157391 \imo $&$ 1.070195-0.157365 \imo $\\
   \hline
$ 0.9 $&$ 1.098214-0.147107 \imo $&$ 1.097317-0.147135 \imo $&$ 1.098209-0.147120 \imo $\\
   \hline
$ 1.0 $&$ 1.129874-0.135250 \imo $&$ 1.128958-0.135270 \imo $&$ 1.129869-0.135263 \imo $\\
   \hline
$ 1.1 $&$ 1.165288-0.121561 \imo $&$ 1.164312-0.121595 \imo $&$ 1.165281-0.121568 \imo $\\
   \hline
$ 1.2 $&$ 1.204540-0.105735 \imo $&$ 1.203323-0.105924 \imo $&$ 1.204472-0.105675 \imo $\\
  \hline
$ 1.3 $&$ 1.247636-0.087381 \imo $&$ 1.245293-0.088836 \imo $&$ 1.244991-0.085730 \imo $\\
   \hline
$ 1.4 $&$ 1.294443 - 0.066050\imo $&$ 1.284790-0.086795 \imo $&$ 0.279216+0.552669 \imo $\\
   \hline
$ 1.45 $&$ 1.319148-0.054135\imo $&$ 1.305462-0.283937 \imo $&$ 1.018660 + 28.96309 \imo $\\
   \hline
\end{tabular}
\caption{Fundamental ($n=0$) quasinormal mode of a massive scalar field in the Schwarzschild background: $\ell=2$ computed by the Frobenius method and WKB formula of 3rd and 6th orders.}
\label{WKBcomp}
\end{table*}

The WKB approach is based on WKB expansion of the effective potential at both infinities (the event horizon and spacial infinity) which are matched with the Taylor expansion near the peak of the effective potential. Therefore, the WKB approach in this form implies existence of the two turning points and monotonic decay of the effective potentials along both
$$
	\frac{i Q_{0}}{\sqrt{2 Q_{0}''}} - \sum_{i=2}^{i=p}
		\Lambda_{i} = n+\frac{1}{2},\qquad n=0,1,2\ldots,
$$
where the correction terms $\Lambda_{i}$ were obtained in \cite{Iyer:1986np} for different orders.  Here $Q_{0}^{i}$ means the i-th derivative of $Q$ at its maximum with respect to the tortoise coordinate $r^\star$, and $n$ labels the overtones.

Here, we shall answer the following question: can the Will-Schutz-Iyer WKB formula \cite{Iyer:1986np}, deduced under assumption of the two turning points, be applied for calculating quasinormal frequencies of massive fields, which imply three turning points.  In order to avoid  the problem connected to the dependence of the effective potential on $\omega$ (which takes place for massive fermions), we shall consider the simplest massive scalar field for illustration. The WKB method works well for $\ell > n $, so that it is evident that for the lowest multipole number $\ell=0$ WKB formula of \cite{Iyer:1986np} should not produce reliable results, which is also confirmed by our comparison with the accurate data. Therefore, we shall consider the case of moderate $\ell$ and $n < \ell$ for which the WKB data of massless fields are known to be reasonably accurate. Comparison of quasinormal frequencies found by the 3rd and 6th order WKB formula with the accurate data obtained via the Frobenius method \cite{Konoplya:2004wg} are shown in table~\ref{WKBcomp}. There, one can see that for small and moderate values of $\mu$ the results found by the WKB formula is in a very good agreement with the accurate values, while for sufficiently large $\mu$, when the effective potential approaches to the transition from the potential barrier to monotonic behavior, WKB formula produces an error which is bigger than the effect of non-zero mass $\mu$.  This cannot be treated by adding more WKB orders, as can be noticed from the comparison of 3rd and 6th WKB orders in the last two rows in Table~\ref{WKBcomp}. While for small and moderate $\mu$ adding more WKB orders gives better accuracy, starting from $\mu r_+ \approx 1.3$ higher WKB orders give bigger error than the lower ones. That is not unexpected as the convergence of the WKB is guaranteed only asymptotically, but not in each order. Thus, we conclude that although WKB formula of \cite{Iyer:1986np} can be applied to the case of massive fields \cite{Simone:1991wn}, it should be used with a great care in the regime of not small masses $\mu$ and it definitely cannot be effectively used for detecting long-lived modes. Probably, a method which takes into consideration all the three turning points, similar to the one suggested in \cite{Galtsov:1991nwq}, could provide better accuracy.


\begin{thebibliography}{80}

\bibitem{TheLIGOScientific:2016src}
  B.~P.~Abbott {\it et al.} [LIGO Scientific and Virgo Collaborations],
  Phys.\ Rev.\ Lett.\  {\bf 116}, no. 6, 061102 (2016)
  [arXiv:1602.03837 [gr-qc]];
  Phys.\ Rev.\ Lett.\  {\bf 116}, no. 22, 221101 (2016)
  [arXiv:1602.03841 [gr-qc]];
  Phys.\ Rev.\ Lett.\  {\bf 116}, no. 24, 241103 (2016)
  [arXiv:1606.04855 [gr-qc]].
\bibitem{Konoplya:2016pmh}
  R.~Konoplya and A.~Zhidenko,
  Phys.\ Lett.\ B {\bf 756}, 350 (2016)
  [arXiv:1602.04738 [gr-qc]];
  JCAP {\bf 1612}, no. 12, 043 (2016)
  [arXiv:1606.00517 [gr-qc]];
  N.~Yunes, K.~Yagi and F.~Pretorius,
  Phys.\ Rev.\ D {\bf 94}, no. 8, 084002 (2016)
  [arXiv:1603.08955 [gr-qc]].

\bibitem{Kerr:1963ud}
  R.~P.~Kerr,
  Phys.\ Rev.\ Lett.\  {\bf 11}, 237 (1963).

\bibitem{reviews}
  R.~A.~Konoplya and A.~Zhidenko,
  Rev.\ Mod.\ Phys.\  {\bf 83}, 793 (2011)
  [arXiv:1102.4014 [gr-qc]];
  E.~Berti, V.~Cardoso and A.~O.~Starinets,
  Class.\ Quant.\ Grav.\  {\bf 26}, 163001 (2009)
  [arXiv:0905.2975 [gr-qc]];
  K.~D.~Kokkotas and B.~G.~Schmidt,
  Living Rev.\ Rel.\  {\bf 2}, 2 (1999)
  [gr-qc/9909058].

\bibitem{Chandrasekhar:1976ap}
  S.~Chandrasekhar,
  Proc.\ Roy.\ Soc.\ Lond.\ A {\bf 349}, 571 (1976).

\bibitem{Cho:2003qe}
  H.~T.~Cho,
  Phys.\ Rev.\ D {\bf 68}, 024003 (2003)
  [gr-qc/0303078].

\bibitem{DiracQNMs}
  J.~l.~Jing,
  Phys.\ Rev.\ D {\bf 71}, 124006 (2005)
  [gr-qc/0502023];
  Y.~J.~Wu and Z.~Zhao,
  Phys.\ Rev.\ D {\bf 69}, 084015 (2004);
  H.~T.~Cho, A.~S.~Cornell, J.~Doukas and W.~Naylor,
  Phys.\ Rev.\ D {\bf 75}, 104005 (2007)
  [hep-th/0701193];
  Phys.\ Rev.\ D {\bf 77}, 041502 (2008)
  [arXiv:0710.5267 [hep-th]];
  S.~K.~Chakrabarti,
  Eur.\ Phys.\ J.\ C {\bf 61}, 477 (2009)
  [arXiv:0809.1004 [gr-qc]];
  R.~Becar, P.~A.~Gonzalez and Y.~Vasquez,
  Phys.\ Rev.\ D {\bf 89}, no. 2, 023001 (2014)
  [arXiv:1306.5974 [gr-qc]];
  J.~Li, M.~Hong and K.~Lin,
  Phys.\ Rev.\ D {\bf 88}, 064001 (2013)
  [arXiv:1308.6499 [gr-qc]];
  P.~A.~Gonz\'alez and Y.~V\'asquez,
  Eur.\ Phys.\ J.\ C {\bf 74}, no. 7, 2969 (2014)
  [arXiv:1404.5371 [gr-qc]];
  K.~S.~Gupta, T.~Juri\'c and A.~Samsarov,
  JHEP {\bf 1706}, 107 (2017)
  [arXiv:1703.00514 [hep-th]];
  K.~H.~C.~Castello-Branco, R.~A.~Konoplya and A.~Zhidenko,
  Phys.\ Rev.\ D {\bf 71}, 047502 (2005)
  [hep-th/0411055];
  O.~P.~F.~Piedra, F.~Sosa, J.~L.~Bernal-Castillo and Y.~Jimenez,
  Int.\ J.\ Mod.\ Phys.\ D {\bf 21}, 1250044 (2012)
  [arXiv:1106.3906 [hep-th]];
  M.~Wang, C.~Herdeiro and J.~Jing,
  Phys.\ Rev.\ D {\bf 96}, no. 10, 104035 (2017)
  [arXiv:1710.10461 [gr-qc]].

\bibitem{He:2006jv}
  X.~He and J.~Jing,
  Nucl.\ Phys.\ B {\bf 755}, 313 (2006)
  [gr-qc/0611003].

\bibitem{Iyer:1986np} B.~F.~Schutz and C.~M.~Will Astrophys.\ J.\ Lett {\bf 291} L33 (1985);
  S.~Iyer and C.~M.~Will,
  Phys.\ Rev.\ D {\bf 35}, 3621 (1987);
  R.~A.~Konoplya,
  Phys.\ Rev.\ D {\bf 68}, 024018 (2003)
  [gr-qc/0303052];
  J.\ Phys.\ Stud.\  {\bf 8}, 93 (2004);
  J.~Matyjasek and M.~Opala,
  Phys.\ Rev.\ D {\bf 96}, no. 2, 024011 (2017)
  [arXiv:1704.00361 [gr-qc]].

\bibitem{Rosa:2011my}
  J.~G.~Rosa and S.~R.~Dolan,
  Phys.\ Rev.\ D {\bf 85} (2012) 044043
  [arXiv:1110.4494 [hep-th]].

\bibitem{angulareq}
K.~G.~Suffern, E.~D.~Fackerell and C.~M.~Cosgrove,
J. Math. Phys. {\bf 24}, 1350 (1983).

\bibitem{Dolan:2009kj}
  S.~Dolan and J.~Gair,
  Class.\ Quant.\ Grav.\  {\bf 26}, 175020 (2009)
  [arXiv:0905.2974 [gr-qc]].

\bibitem{Dolan:2015eua}
  S.~R.~Dolan and D.~Dempsey,
  Class.\ Quant.\ Grav.\  {\bf 32}, no. 18, 184001 (2015)
  [arXiv:1504.03190 [gr-qc]].

\bibitem{Huang:2017nho}
  Y.~Huang, D.~J.~Liu, X.~h.~Zhai and X.~z.~Li,
  Phys.\ Rev.\ D {\bf 96}, no. 6, 065002 (2017)
  [arXiv:1708.04761 [gr-qc]].

\bibitem{Ohashi:2004wr}
  A.~Ohashi and M.~a.~Sakagami,
  Class.\ Quant.\ Grav.\  {\bf 21}, 3973 (2004).
  [gr-qc/0407009].

\bibitem{Konoplya:2004wg}
  R.~A.~Konoplya and A.~Zhidenko,
  Phys.\ Lett.\ B {\bf 609}, 377 (2005)
  [gr-qc/0411059].

\bibitem{Konoplya:2005hr}
  R.~A.~Konoplya,
  Phys.\ Rev.\ D {\bf 73}, 024009 (2006)
  [gr-qc/0509026].

\bibitem{quasiresonances}
  R.~A.~Konoplya and A.~Zhidenko,
  Phys.\ Rev.\ D {\bf 73}, 124040 (2006)
  [gr-qc/0605013];
  A.~Zhidenko,
  Phys.\ Rev.\ D {\bf 74}, 064017 (2006)
  [gr-qc/0607133];
  C.~Wu and R.~Xu,
  Eur.\ Phys.\ J.\ C {\bf 75}, no. 8, 391 (2015)
 [arXiv:1507.04911 [gr-qc]];
  B.~Toshmatov and Z.~Stuchlik,
  Eur.\ Phys.\ J.\ Plus {\bf 132}, no. 7, 324 (2017)
  [arXiv:1707.07419 [gr-qc]].

\bibitem{Konoplya:2007yy}
  R.~A.~Konoplya and R.~D.~B.~Fontana,
  Phys.\ Lett.\ B {\bf 659}, 375 (2008)
  [arXiv:0707.1156 [hep-th]].

\bibitem{Su:2017fcm}
  D.~Su, C.~T.~M.~Ho, R.~B.~Mann and T.~C.~Ralph,
  Phys.\ Rev.\ D {\bf 96}, no. 6, 065017 (2017)
  [arXiv:1706.09117 [gr-qc]].

\bibitem{Seahra:2004fg}
  S.~S.~Seahra, C.~Clarkson and R.~Maartens,
  Phys.\ Rev.\ Lett.\  {\bf 94}, 121302 (2005)
  [gr-qc/0408032].

\bibitem{Yang:2012pj}
  H.~Yang, F.~Zhang, A.~Zimmerman, D.~A.~Nichols, E.~Berti and Y.~Chen,
  Phys.\ Rev.\ D {\bf 87}, no. 4, 041502 (2013)
  [arXiv:1212.3271 [gr-qc]].

\bibitem{ZDMs}
  S.~Hod,
  Phys.\ Rev.\ D {\bf 78}, 084035 (2008)
 [arXiv:0811.3806 [gr-qc]];
  K.~D.~Kokkotas, R.~A.~Konoplya and A.~Zhidenko,
  Phys.\ Rev.\ D {\bf 92}, no. 6, 064022 (2015)
  [arXiv:1507.05649 [gr-qc]].

\bibitem{Richartz:2017qep}
  M.~Richartz, C.~A.~R.~Herdeiro and E.~Berti,
  Phys.\ Rev.\ D {\bf 96}, no. 4, 044034 (2017)
  [arXiv:1706.01112 [gr-qc]].

\bibitem{Burko:2017eky}
  L.~M.~Burko and G.~Khanna,
  arXiv:1709.10155 [gr-qc].

\bibitem{Yang:2014tla}
  H.~Yang, A.~Zimmerman and L.~Lehner,
  Phys.\ Rev.\ Lett.\  {\bf 114}, 081101 (2015)
  [arXiv:1402.4859 [gr-qc]].

\bibitem{Page:1976jj}
  D.~N.~Page,
  Phys.\ Rev.\ D {\bf 14}, 1509 (1976).

\bibitem{Zhidenko:2009zx}
  A.~Zhidenko,
  arXiv:0903.3555 [gr-qc].

\bibitem{Leaver:1985ax}
 Leaver,~E.~W.,
 Proc.\ Roy.\ Soc.\ Lond.\  A {\bf 402}, 285 (1985).

\bibitem{Konoplya:2007zx}
  R.~A.~Konoplya and A.~Zhidenko,
  Phys.\ Rev.\ D {\bf 76}, no. 8, 084018 (2007)
  Erratum: [Phys.\ Rev.\ D {\bf 90}, no. 2, 029901 (2014)]
  [arXiv:0707.1890 [hep-th]].

\bibitem{Konoplya:2013rxa}
  R.~A.~Konoplya and A.~Zhidenko,
  Phys.\ Rev.\ D {\bf 88}, 024054 (2013)
  [arXiv:1307.1812 [gr-qc]].

\bibitem{Blazquez-Salcedo:2017bld}
  J.~L.~Bl\'azquez-Salcedo and C.~Knoll,
  Phys.\ Rev.\ D {\bf 97}, no. 4, 044020 (2018)
  [arXiv:1709.07864 [gr-qc]].

\bibitem{Simone:1991wn}
  L.~E.~Simone and C.~M.~Will,
  Class.\ Quant.\ Grav.\  {\bf 9} (1992) 963;
  R.~A.~Konoplya,
  Phys.\ Lett.\ B {\bf 550}, 117 (2002)
  [gr-qc/0210105].


\bibitem{Galtsov:1991nwq}
  D.~V.~Gal'tsov and A.~A.~Matiukhin,
  Class.\ Quant.\ Grav.\  {\bf 9}, 2039 (1992).

\end{thebibliography}
\end{document}